\documentclass[a4paper,12pt]{article}

\usepackage{amsmath}
\usepackage{amssymb}
\usepackage{amsthm}

\newcommand{\prt}{\partial}

\def\F{{\mathcal F}}
\def\D{{\mathcal D}}

\def\B{{\tilde B}}

\def\L{{\tilde L}}

\def\ni{{\noindent}}

\def\be{\begin{equation}}
\def\ee{\end{equation}}

\newtheorem{theorem}{Theorem}

\theoremstyle{definition}

\begin{document}
\bibliographystyle{plain}

\title{{\Large\bf  Hard edge for beta-ensembles and Painlev\'e III.}
\author{Igor Rumanov\footnote{e-mail: igor.rumanov@colorado.edu} \\ 
{\small CU Boulder, Boulder, CO}}}

\maketitle

\bigskip

\begin{abstract}
Starting from the diffusion equation at beta random matrix hard edge obtained by Ramirez and Rider (2008), we study the question of its relation with Lax pairs for Painleve III. The results are in many respects similar to the ones found for soft edge by Bloemendal and Virag (2010). In particular, the values beta = 2 and 4 (but not beta = 1) allow for a simple connection with Painlev\'e III solutions and Lax pairs. However, there is an additional surprise for a special relation of parameters where a simple solution of the diffusion equation can be obtained, which is a one-parameter generalization of Gumbel distribution. Our considerations can be extended to the other Painleve equations since the corresponding diffusions  are in fact known as nonstationary (imaginary time) Schr\"odinger equations for quantum Painlev\'e Hamiltonians. We also track the hard-to-soft edge limit transition in terms of our Lax pairs.
\end{abstract}

\newpage

\section{Introduction}

Beta-ensembles of random matrices were originally defined by their eigenvalue probability distribution depending on real nonnegative parameter denoted $\beta$ by Dyson~\cite{DyBeta} who first introduced the probability density function (p.d.f.),

$$
P_{\beta}(l_1, \dots, l_n) = \frac{1}{Z_{\beta}}\cdot\prod_{i<j}|l_i-l_j|^{\beta}\cdot\prod_{k=1}^ne^{-V(l_k)},   \eqno(1.1)
$$

\ni where $V(x)$ is called the potential and $Z_{\beta}$ is a normalization constant. This expression is the joint eigenvalue p.d.f. for three important invariant ensembles of random matrices, whose entries have probability distribution symmetric with respect to the real orthogonal group ($\beta=1$), complex unitary group ($\beta=2$) and symplectic group ($\beta=4$). For the other values of $\beta$ there had been no known random matrix model realizing the above eigenvalue distribution until 2002. Then Dumitriu and Edelman~\cite{DE02} discovered tridiagonal random matrix ensembles which lead to (1.1) for $V(x)=x^2$ (Gaussian $\beta$ ensemble) and $V(x) = -x+\text{const.}\cdot\ln x$ (Laguerre (or Wishart) $\beta$ ensemble). This discovery spurred a lot of further research on $\beta$-ensembles and their large size $n$ limits. New and different matrix models for $\beta$-ensembles, e.g.~invariant rather than tridiagonal~\cite{ABG12}, have been also constructed. Even without the corresponding matrix ensembles, the eigenvalue distributions (1.1) have physical significance as description of Coulomb gas(fluid) restricted to a line, which has application in problems of conductivity and other transport phenomena in various disordered systems, see e.g.~\cite{F2010} and references therein. 
\par Their theoretical importance is also due to the relation with conformal field theory exposed by Virasoro subalgebra constraints they satisfy~\cite{Awata, AvM7, IR1}. This relation implies that for general $\beta$ these eigenvalue distributions and matrix ensembles are natural subject of quantum integrable systems theory~\cite{BLZ, ChEyM, F2010}. Indeed, recently their direct connection with (imaginary time) canonically quantized Painlev\'e Hamiltonians was shown for some special potentials~\cite{Nag11}. There certain Fokker-Planck (FP) (or diffusion-drift) equations reappeared which had been found before by completely different, probabilistic, approach in~\cite{BV} based on earlier work~\cite{RRV} for random $\beta$-matrix soft edge and in~\cite{RR08, RR12} for the hard edge, the two important scaling limits for large $n$. 
\par In the last references, the tridiagonal matrices of~\cite{DE02} and their spiked\footnote{In statistics, a Wishart sample covariance matrix is called spiked if the underlying population covariance matrix has a single eigenvalue different from unity} modifications were the starting point to rigorously derive random Schr\"odinger equations in the large $n$ limit, conjectured earlier in~\cite{EdSut}. Then they were reduced to first-order stochastic ODEs (of Langevin type though nonlinear) by the Riccati transformation first used likely by~\cite{Hal65}. Then the corresponding probability distributions satisfied the FP equations featuring the quantum Painlev\'e II (for the soft edge) and quantum Painlev\'e III (for the hard edge) Hamiltonians. In~\cite{BV} the soft edge FP equation 

$$
(\prt_t + \frac{2}{\beta}\prt_{xx} + (t-x^2)\prt_x)\F^S_{\beta} = 0,   \eqno(1.2)
$$

\ni where $t$ is the scaled edge of the eigenvalue spectrum and $x$ is the scaled spike parameter, and its bounded solution $\F^S_{\beta}$ describing the largest eigenvalue probability for critically spiked large $n$ Wishart $\beta$-ensemble was shown to be related to Baik-Rains Lax pair~\cite{BR01} for (classical) Painlev\'e II equation, however, only for $\beta=2$ and $4$. This pair,

$$
\prt_x\left(\begin{array}{c}f \\ g \end{array}\right) = \left(\begin{array}{cc} q^2  & -(qx+q') \\ -qx+q' & x^2-t-q^2  \end{array}\right)\left(\begin{array}{c}f \\ g \end{array}\right),   
$$

$$
\prt_t\left(\begin{array}{c}f \\ g \end{array}\right) = \left(\begin{array}{cc} 0  & q \\ q & -x  \end{array}\right)\left(\begin{array}{c}f \\ g \end{array}\right),   \eqno(1.3)
$$

\ni is a modification of Flaschka-Newell~\cite{FN80} Lax pair found earlier to be relevant for the critical regime in the Tracy-Widom-to-Gaussian phase transition for the largest eigenvalue distribution of $n\to\infty$ complex Wishart ($\beta=2$) ensemble~\cite{BBP}. It was found in~\cite{BV} that 

$$
F^S_2(t, x) = F_2(t)f(t,x),  \ \ \ \ \ F^S_4(t,x) = F_2^{1/2}(t)(\varphi(t)f(t,x) + \gamma(t)g(t,x))   \eqno(1.4)
$$

\ni where $F_2(t)$ is the Tracy-Widom distribution~\cite{TW-Airy}, and, thus, for these special $\beta$ only, the solution of the FP equation is proportional to either a component of the Lax pair eigenvector or a linear combination of its components. This turned out not so for the other values of $\beta$ and e.g., surprisingly, for $\beta=1$, where results could have been expected similar to $\beta=4$ case, which is related to it by duality $\beta \leftrightarrow 4/\beta$ playing a big role in many aspects of $\beta$-matrix models, see e.g.~\cite{TW96, Der09, MuWal02}. The part of the operator on the left-hand side of eq.~(1.2) with $x$-derivatives is nothing but the canonically quantized Painlev\'e II Hamiltonian which appeared as such in~\cite{Nag11} (up to the change of sign of $t$).
\par {\bf Remark.} Just recently a nice formula for the spiked general $\beta$ finite $n$ Wishart matrix p.d.f. was obtained by Forrester~\cite{ForBeta11} based on several earlier results for special cases: 

$$
P_{\beta, a, \delta}(l_1, \dots, l_n) = \frac{1}{Z_{\beta, a, \delta}}\cdot\prod_{i<j}|l_i-l_j|^{\beta}\cdot\prod_{k=1}^nl_k^{(a+1)\beta/2 - 1}e^{-\beta l_k/2}\int_{-\infty}^{+\infty}e^{it}\prod_{k=1}^n\left(it - \frac{\delta-1}{2\delta}l_k\right)^{-\beta/2}dt,
$$

\ni where $\delta$ is the (non-scaled) spike parameter and in statistical context $a$ is the integer difference of dimensions of Gaussian rectangular $n\times m$ sample matrix $X$ whose covariance matrix $XX^{\dagger}$ is then $n\times n$ Wishart matrix. However, in general, the only restriction on $a$ is $a>-1$ needed for convergence of the integrals over the above density. The hard edge limit arises here when $n\to \infty$, $n\delta \to x$, and the scaled spike parameter $x$ is fixed. \\

\par Another FP equation describes the hard edge large $n$ limit of $\beta$-ensembles, e.g.~of (spiked) Wishart(Laguerre) ensemble, for the smallest eigenvalues restricted by hard wall of positivity constraint. This was derived in~\cite{RR08, RR12} for different boundary conditions depending on whether the spiked or non-spiked case is studied. The equation is

$$
\left(\prt_{\mu} + \frac{2}{\beta}x^2\prt_{xx} + \left((a+\frac{2}{\beta})x - x^2 - e^{-\mu}\right)\prt_x\right)\F^H(x, \mu) = 0,    \eqno(1.5)
$$

\ni and for the spiked case the boundary conditions are

$$
\F^H(-\infty, x) = 0,   \ \ \ \F^H(\infty, x) = 1,  \ \ \ \F^H(\mu, 0) = 0.   \eqno(1.6)
$$

\ni The operator on the left-hand side of eq.~(1.5) is this time $\prt_{\mu}$ plus the canonically quantized Painlev\'e III (PIII) Hamiltonian which appeared as such in~\cite{Nag11}, up to the change of variable $e^{-\mu} \to -t$. The FP equation (1.5) is the main object of study here, with the goal of exploring the possible connections with classical integrability and comparing the results with that of soft edge~\cite{BV}. Motivated also by~\cite{Nag11}, we look for Lax pairs which give Painlev\'e III equation as consistency condition such that the function $\F^H$ is a component of their eigenvector. The dependence of such Lax pairs on spectral parameter, whose role is played by the spiking parameter $x$ here, is rational and known in general for all Painlev\'e equations~\cite{JM81}. With this example of PIII, we develop a general method of finding such quantum-classical correspondences, which should work equally well e.g.~for the other Painlev\'e equations and more broadly, a conceptual discussion of this approach to appear~\cite{BetaPhil}. Here, essentially, the compatibility of FP equation (1.5) and a Lax pair for classical PIII, which brings additional algebraic constraints imposed on the elements of the Lax matrices, which not always can be successfully resolved to find nontrivial solution. When they can, some of them turn out to be first integrals for the ODEs involved which in effect prevents the whole system of ODEs and algebraic constraints from being overdetermined.
\par One should mention also the studies of multi-spiked Wishart ensembles and their soft edge~\cite{BBP, BV2} and hard edge~\cite{RR12} where multidimensional generalizations of the above FP equations emerge. Here again connections with quantum integrability are implied for general $\beta$, see e.g.~\cite{Nag11} and references therein, and explicit determinantal formula for $\beta=2$ distribution is known~\cite{BaikP06} in terms of the function $f$ from eq.~(1.3). But here we consider only the one-spike case.

\section{Main results}

We set up and solve the problem of finding possible classical Lax pairs for Painlev\'e III equation, such that their first component satisfies eq.~(1.5). The results obtained are in many respects similar to the ones for soft edge case~\cite{BV}. Again, two values $\beta=2,4$, but not $\beta=1$, turn out to be the special ones under which there are the sought Lax pair related to Painlev\'e III. Besides these two, however, here appears another special value, more precisely, special relation:

$$
\frac{\beta}{2} + \frac{2}{\beta} = 2 - a,   \eqno(2.1)
$$
  
\ni which admits a special unusual Lax pair depending on two arbitrary functions of $\mu$. Actually we formulate our results in different variables. We change the independent variable $\mu$ for $t = e^{\mu}$, multiply eq.~(1.5) by $\kappa = \beta/2$, rescale variables as $\kappa x \to x$, $t \to \kappa^2t$ and redenote $a+1/\kappa \to a$, and arrive at 

$$
(\kappa t\prt_t + x^2\prt_{xx} + (ax - x^2 - 1/t)\prt_x)\F^H(t,x) = 0,    \eqno(2.2) 
$$

\noindent with the corresponding boundary conditions derived from eq.~(1.6), 

$$
\F^H(0, x) = 0,   \ \ \ \F^H(+\infty, x) = 1,  \ \ \ \F^H(t, 0) = 0.   \eqno(2.3)
$$

\ni Now we can write out the main outcomes explicitly.  Consider Lax pairs of the form:

$$
\prt_x\left(\begin{array}{c}\F^H \\ G \end{array}\right) = L\left(\begin{array}{c}\F^H \\ G \end{array}\right),    \eqno(2.4)
$$

$$
\prt_t\left(\begin{array}{c}\F^H \\ G \end{array}\right) = B\left(\begin{array}{c}\F^H \\ G \end{array}\right),  \eqno(2.5)
$$

\ni Then we obtain the following:

\begin{theorem}
If $\beta=2$, the solution $\F^H(t,x)$ of eq.~(2.2) is the eigenvector first component of the Lax pair of the form (2.4), (2.5) with


$$
L =  \left(\begin{array}{cc} \frac{q^2}{tx^2}  & \frac{\phi}{x} + \frac{ty\phi}{x^2} \\ \frac{q^2h}{t^2y\phi}\frac{1}{x} - \frac{q^2(q^2 - 1)}{t^3y\phi}\frac{1}{x^2} & \frac{1-q^2}{tx^2} + \frac{1-a}{x} + 1  \end{array}\right),
$$


$$
B = \left(\frac{q^2}{t^2} - \frac{q^2h}{t^3y}\right)\cdot I + \frac{1}{x}\left(\begin{array}{cc}  \frac{q^2}{t^2} & y\phi \\ -\frac{q^2(q^2 - 1)}{t^4y\phi} & \frac{1-q^2}{t^2}  \end{array}\right),    \eqno(2.6a)
$$

\ni where function $y(t)=\phi'(t)/\phi$ is a solution of the following Painlev\'e III equation (`prime' here and further on means derivative w.r.t.~$t$), 

$$
y'' = \frac{(y')^2}{y} - \frac{1}{t}y' + y^3 + \frac{a}{t}y^2 - \frac{(a-1)}{t^4} - \frac{1}{t^6y},  \eqno(2.7a)
$$

\ni functions $q(t)$ and $y(t)$ satisfy the system of ODEs, 

$$
t^3y' = 2q^2 - 1 - t^3y^2 - (a+1)t^2y,  \eqno(2.8a)
$$

$$
tq' = \frac{q(q^2-1)}{t^2y} - \frac{(a-1)}{2}q,   \eqno(2.9a)
$$



\ni  and function $h(t)$ is

$$
h =  \frac{q^2-1}{t^2y} - (a-1).    \eqno(2.10a)
$$



If $\beta=2$, the solution $\F^H(t,x)$ of eq.~(2.2) is also the eigenvector first component of the Lax pair of the form (2.4), (2.5) with

$$
L =  \left(\begin{array}{cc} q^2  & \frac{\phi}{x} - t^2y\phi \\ \frac{q^2h}{t^2y\phi}\frac{1}{x} + \frac{q^2(q^2 - 1)}{t^2y\phi} & \frac{1}{tx^2} - \frac{a}{x} + 1 - q^2 \end{array}\right),
$$

$$
B = \left(\frac{q^2}{t^2} - \frac{q^2h}{t^3y}\right)\cdot I + \left(\begin{array}{cc}  0 & \frac{\phi}{t} \\ \frac{q^2h}{t^3y\phi} & \frac{1}{t^2x} - \frac{a}{t}  \end{array}\right),    \eqno(2.6b)
$$

\ni where function $y(t)=\phi'(t)/\phi$ is a solution of the following Painlev\'e III equation 

$$
y'' = \frac{(y')^2}{y} - \frac{1}{t}y' + y^3 - \frac{a-1}{t}y^2 + \frac{a}{t^4} - \frac{1}{t^6y},  \eqno(2.7b)
$$

\ni functions $q(t)$ and $y(t)$ satisfy the system of ODEs, 

$$
t^3y' = 2q^2 - 1 - t^3y^2 + (a-2)t^2y,  \eqno(2.8b)
$$

$$
tq' = \frac{q(q^2-1)}{t^2y} + \frac{a}{2}q,   \eqno(2.9b)
$$

\ni  and this time function $h(t)$ is

$$
h =  \frac{q^2-1}{t^2y} + a.    \eqno(2.10b)
$$

\end{theorem}

{\bf Remark.} Function $q$ from the Lax pair (2.6b) itself satisfies eq.~(2.12) below, a variant of Tracy-Widom equation for $\beta=2$ hard edge from~\cite{TW-Bes}. If we denote it $q(t;a)$ and denote $y(t;a)$ the solution of eq.~(2.7b), then functions $q$ and $y$ from the Lax pair (2.6a) are nothing but $q(t; 1-a)$ and $y(t; 1-a)$, respectively. Nontrivial consistency conditions between the pairs (2.6a) and (2.6b) arise from the fact that they involve the same function $\F^H$ in the eigenvectors. They lead to the following two algebraic identities, which can be directly verified from ODEs satisfied by $q$ and $y$: 

$$
y(t;a)y(t;1-a) = -\frac{1}{t^3},  \ \ \ \ \ \frac{q^2(t;a)}{y(t;a)} + \frac{q^2(t;1-a)}{y(t;1-a)} = 0.
$$

\bigskip

The function $q(t;a)$ also relates the solutions at $\beta=2$ and $\beta=4$. One has

\begin{theorem}
If $\beta=4$, the solution $\F^H(t,x)$ of eq.~(2.2) is the eigenvector first component of the Lax pair of the form (2.4), (2.5) with

$$
L = \left(\begin{array}{cc} \frac{1+q}{2tx^2}-\frac{a}{2x}+\frac{1-q}{2} & \frac{S_+}{2tx^2}+\frac{2tq'+a}{2(q^2-1)}\frac{S_+}{x}+\frac{S_+}{2} \\ \frac{S_-}{2tx^2}-\frac{2tq'-a}{2(q^2-1)}\frac{S_-}{x}+\frac{S_-}{2} & \frac{1-q}{2tx^2}-\frac{a}{2x}+\frac{1+q}{2}  \end{array}\right),
$$

$$
B = h_0(t)\cdot I + \left(\begin{array}{cc} \frac{1+q}{2t^2x} & \frac{S_+}{2t^2x}+\frac{(2tq'+a)S_+}{2t(q^2-1)} \\ \frac{S_-}{2t^2x}-\frac{(2tq'-a)S_-}{2t(q^2-1)} & \frac{1-q}{2t^2x}  \end{array}\right),    \eqno(2.11)
$$

\ni where function $q(t)$ satisfies the following Painlev\'e III-related equation, a variant of which first appeared in~\cite{TW-Bes} (but there $\beta=2$):

$$
t(q^2-1)(tq')' = q(tq')^2 + \frac{1}{t}q^3(q^2-2) + \left(\frac{1}{t} - \frac{a^2}{4}\right)q,   \eqno(2.12)
$$

\ni functions $S_+(t)$ and $S_-(t)$ are determined by 

$$
\frac{S_+'}{S_+} = \frac{qq'}{q^2-1} + \frac{aq}{2t(q^2-1)}, \ \ \ \ \frac{S_-'}{S_-} = \frac{qq'}{q^2-1} - \frac{aq}{2t(q^2-1)},  \ \ \ \ S_+S_- = 1-q^2,   \eqno(2.13)
$$

\ni and function $h(t)$ is

$$
h_0 = \frac{1}{2t}\left(-\frac{(2tq')^2-a^2}{4(q^2-1)} + \frac{q^2}{t} + \frac{a(a-2)}{4}\right).   \eqno(2.14)
$$

\end{theorem}

\begin{theorem}
If eq.~(2.1) holds, the solution $\F^H(t,x)$ of eq.~(2.2) is the eigenvector first component of the Lax pair of the form (2.4), (2.5) with

$$
L = \left(\begin{array}{cc} \frac{1+r}{2tx^2} & \frac{\phi}{tx^2} \\ -\frac{r^2-1}{4t\phi x^2} & \frac{1-r}{2tx^2} \end{array}\right),
$$

$$
B = \left(\begin{array}{cc} \frac{1+r}{2t^2x} + \frac{1+r}{2\kappa t^2} & \frac{\phi}{t^2x}+\frac{\phi}{\kappa t^2} \\ -\frac{r^2-1}{4t^2\phi x}-\frac{r^2-1}{4\kappa t^2\phi} - \frac{r'}{2\phi} & \frac{1-r}{2t^2x} + \frac{1-r}{2\kappa t^2} - \frac{\phi'}{\phi} \end{array}\right).   \eqno(2.15)
$$

\ni For {\it any} functions $r(t)$ and $\phi(t)$, the Lax equations are satisfied {\it identically} with this pair. The FP eq.~(2.2) is satisfied by the first component $\F^H$ of the eigenvector only if eq.~(2.1) is imposed. These conditions lead, however, to a simple explicit distribution function $\F^H(t,x)$, unrelated to Painlev\'e III,

$$
\F^H(t, x) = \exp{\left(-\frac{x+\kappa}{\kappa tx}\right)}.   \eqno(2.16)
$$

\ni Thus, here only the partner of $\F^H$, $G(t,x)$, depends on the two arbitrary functions. It is given by 

$$
G(t,x) = \frac{1-r}{2\phi}\F^H.   \eqno(2.17)
$$

\end{theorem}

\ni Only the above options for $\beta$ give the possibility for $\F^H(t,x)$ of eq.~(2.2) to be related with classical Lax pair in this simple way.
\par It is well known that when $a$ becomes large, the smallest eigenvalues of random matrices separate from the hard wall at zero and the hard edge of the spectrum turns into a soft edge~\cite{BorFor}, see also~\cite{ForPainl00, F2010, RR08}. Then, under proper scaling,

$$
x \to \alpha(1+\alpha^{-1/3}x),  \ \ \ \ \ t \to \alpha^{-2}(1+\alpha^{-2/3}t), \ \ \ \ \ \alpha = \frac{a}{2},   \eqno(2.18)  
$$

\ni the solution $\F^H$ of eqs.~(2.2), (2.3) transforms~\cite{RR08, RR12} into the bounded solution $\F^S$ of eq.~(1.2) after changing $\kappa^2t\to t$ and $\kappa x \to x$ there. It is interesting also to follow this transition in terms of the above Lax pairs (2.6a), (2.6b) and (2.11). Doing this, we obtain the next

\begin{theorem}
Under the scaling (2.18), when $\alpha\to \infty$, the Lax pairs (2.6a) and (2.6b) transform into the pair

$$
\prt_x\left(\begin{array}{c}\F^S \\ G^S \end{array}\right) = \left(\begin{array}{cc} q^2  & -(qx+q') \\ -qx+q' & x^2-t-q^2  \end{array}\right)\left(\begin{array}{c}\F^S \\ G^S \end{array}\right),   
$$

$$
\prt_t\left(\begin{array}{c}\F^S \\ G^S \end{array}\right) = \left(\begin{array}{cc} u  & q \\ q & u-x  \end{array}\right)\left(\begin{array}{c}\F^S \\ G^S \end{array}\right),   \eqno(2.19)
$$

\ni where $q(t)$ is the Hastings-McLeod solution of Painlev\'e II equation,

$$
q'' = tq + 2q^3,   \eqno(2.20)
$$

\ni and function $u(t)$ satisfies

$$
u = (q')^2-tq^2-q^4, \ \ \ \ \ u' = -q^2.   \eqno(2.21)
$$

\ni The components of the eigenvector $(\F^H, G)$ of (2.6a) scale as

$$
\F^H \to \alpha^{-2/3}\F_S, \ \ \ \ \ G \to e^{\alpha^{1/3}t}G^S,   \eqno(2.22a)
$$

\ni while that of (2.6b) scale as

$$
\F^H \to \alpha^{-2/3}\F_S, \ \ \ \ \ G \to e^{\alpha^{-1/3}t}G^S,   \eqno(2.22b)
$$

\ni Pair (2.19) is almost the same as (1.3), it becomes exactly (1.3) after changing the eigenvector as $(\F^S, G^S) \to F_2(t)(f, g)$. Then $u$ goes away from matrix $B$ in (2.19) since $u$ is also the logarithmic derivative of Tracy-Widom distribution $F_2(t)$: $F_2'(t) = uF_2$.

\end{theorem}

\begin{theorem}
Under the scaling (2.18), when $\alpha\to \infty$, the Lax pair (2.11) transforms into the pair

$$
L \to \alpha^{-2/3} \left(\begin{array}{cc} \frac{x^2-t}{2}-qx  & \left(\frac{x^2-t}{2} - q' - q^2\right)S_+ \\ \left(\frac{x^2-t}{2} + q' - q^2\right)S_- & \frac{x^2-t}{2}+qx   \end{array}\right),   
$$

$$
B \to \alpha^{8/3} \left(\begin{array}{cc} -\frac{x-u-q}{2}  & -\frac{S_+x}{2} \\ -\frac{xS_-}{2} & -\frac{x-u+q}{2}  \end{array}\right),   \eqno(2.23)   
$$

\ni where $q$ is the same Hastings-McLeod solution of eq.~(2.20), the functions $S_+, S_-$ satisfy the relations

$$
S_+' = -qS_+,  \ \ \ \ \ S_+S_- =1,   \eqno(2.24)   
$$

\ni and $u$ is the same as in (2.21). The eigenvector components of (2.11) both rescale as

$$
(\F^H, G) \to \alpha^{-2/3}(\F^S, G^S).    \eqno(2.25)  
$$

\end{theorem}

\ni Here our function $S_+(t)$ is seen to be the function denoted $E^{-1}(x)$ in~\cite{BV} (our $t$ here is their $x$); in our notation, $\varphi(t) = (S_+^{1/2}+S_-^{1/2})/2$, $\gamma(t) = (S_+^{1/2}-S_-^{1/2})/2$, see eq.~(1.4). The representation (2.23) for $\beta=4$ case is still different from the representation in~\cite{BV}, here $\F^S$ and $G^S$ are linear combinations of $f$ and $g$ from (1.3).
\par The rest of the paper is mostly dedicated to the derivation of the above results. In section 3 we obtain the general set of ODEs and constraints relating the solution of FP equation (2.2) with a Lax pair of the form (2.4), (2.5). Then we consider different possible ways of constraint resolution in section 4. In section 5 we derive Painlev\'e III (PIII) eqs. ~(2.7a), (2.7b) and Tracy-Widom eq.~(2.12) from our equations. We then obtain the Lax pairs (2.6a), (2.6b) and (2.11) as well as Theorem 3 in section 6. In section 7 we consider the hard-to-soft edge limit transition and obtain Theorems 4 and 5. Some conclusions are presented in the last section 8.

\bigskip

{\bf\large Acknowledgments.} I would like to thank B. Rider for multiple useful discussions. The financial support by NSF grant DMS-0645756 is gratefully acknowledged. Also I would like to thank MSRI at Berkeley, CA, for support during my stay there in Fall 2010 there as a postdoctoral fellow in Special Program on Random Matrices and Random Processes, where my interest in the current problem was initiated.  

\section{FP equation for the hard edge and Lax pair for PIII}


 A Lax pair for Painlev\'e III equation can be always found in the form such that



$$
L = \frac{L_2(t)}{x^2} + \frac{L_1(t)}{x} + L_0(t),   \eqno(3.1) 
$$

\ni and 

$$
B = B_1(t)x + B_0(t) + \frac{B_{-1}(t)}{x},   \eqno(3.2) 
$$

\ni as follows from singularity theory for ordinary differential equations (ODEs), see~\cite{JM81}. As follows from eq.~(2.4),

$$
\prt_{xx}\left(\begin{array}{c}\F^H \\ G \end{array}\right) = (\prt_xL + L^2)\left(\begin{array}{c}\F^H \\ G \end{array}\right),    \eqno(3.3) 
$$

\ni and, using eq.~(3.1), one finds that

$$
\prt_xL + L^2 = \frac{L_2^2}{x^4} + \frac{\{L_2, L_1\} - 2L_2}{x^3} + \frac{L_1^2 + \{L_2, L_0\} - L_1}{x^2} + \frac{\{L_1, L_0\}}{x} + L_0^2.   \eqno(3.4) 
$$

\ni From eq.~(3.1) one also finds

$$
(ax-x^2-1/t)L = -\frac{L_2/t}{x^2} + \frac{aL_2-L_1/t}{x} + aL_1 - L_2 - L_0/t + (aL_0-L_1)x - L_0x^2.   \eqno(3.5) 
$$

\ni Therefore the action of the hard edge FP operator from eq.~(2.2) on the eigenvector of sought pair (2.4), (2.5) gives

$$
(\kappa t\prt_t + x^2\prt_{xx} + (ax - x^2 - 1/t)\prt_x)\left(\begin{array}{c}\F^H \\ G \end{array}\right) = (\kappa tB + x^2(\prt_xL + L^2) + (ax-x^2-1/t)L)\left(\begin{array}{c}\F^H \\ G \end{array}\right) = 
$$

$$
= \left(\frac{L_2^2-L_2/t}{x^2} + \frac{\kappa tB_{-1} + \{L_2, L_1\} + (a-2)L_2 - L_1/t}{x} + \kappa tB_0 + L_1^2 + \{L_2, L_0\} + \right. 
$$

$$
\left. + (a-1)L_1 - L_2 - \frac{L_0}{t} + (\kappa tB_1 + \{L_1, L_0\} + aL_0-L_1)x + (L_0^2-L_0)x^2\right)\left(\begin{array}{c}\F^H \\ G \end{array}\right).   \eqno(3.6) 
$$

\ni Let us denote the matrix on the r.h.s. of eq.~(3.6) as $\D_H(L, B)$, then, for $\F^H$ to be a solution of eq.~(2.2), we must have

$$
\D_H(L, B)\left(\begin{array}{c}\F^H \\ G \end{array}\right) = \left(\begin{array}{cc} 0 & 0 \\ \D_{21} & \D_{22} \end{array}\right)\left(\begin{array}{c}\F^H \\ G \end{array}\right) = \left(a_d(x,t)(I - \sigma_3) + a_a\sigma_-\right)\left(\begin{array}{c}\F^H \\ G \end{array}\right),   \eqno(3.7) 
$$

\noindent where, according to the powers of $x$ present in $\D_H(L, B)$,

$$
a_d(x, t) = \sum_{k=-2}^2d_k(t)x^k, \ \ \ \ \ a_a(x, t) = \sum_{k=-2}^2a_k(t)x^k,   \eqno(3.8) 
$$

\ni and the functions $d_k(t)$ and $a_k(t)$ are arbitrary so far.
\par Besides the FP equation (3.6), which will give us some constraints, we have the Lax (or, more precisely, the zero curvature) equation:

$$
\prt_tL = \prt_xB + [B, L],   \eqno(3.9) 
$$

\ni which will give us the set of ODEs on the components of $L$ and some more constraints. Since, by (3.1) and (3.2),

$$
[B, L] = \frac{[B_{-1}, L_2]}{x^3} + \frac{[B_{-1}, L_1] + [B_0, L_2]}{x^2} + \frac{[B_{-1}, L_0] + [B_0, L_1] + [B_1, L_2]}{x} + 
$$

$$
+[B_0, L_0] + [B_1, L_1] + [B_1, L_0]x,   \eqno(3.10) 
$$

\ni one obtains the following five matrix equations for the five powers of $x$ present in (3.9):

$$
x^{-3}: [B_{-1}, L_2] = 0,   \eqno(3.11) 
$$

$$
x^{-2}: \prt_tL_2 = -B_{-1} + [B_{-1}, L_1] + [B_0, L_2],   \eqno(3.12) 
$$

$$
x^{-1}: \prt_tL_1 = [B_{-1}, L_0] + [B_0, L_1] + [B_1, L_2],   \eqno(3.13) 
$$

$$
x^{0}: \prt_tL_0 = B_1 + [B_0, L_0] + [B_1, L_1],   \eqno(3.14) 
$$

$$
x^{1}: [B_1, L_0] = 0.   \eqno(3.15) 
$$

\ni Let us separate the trace parts of matrices $L$ and $B$, i.e. write $L= L_t\cdot I + \L$, $B = B_t\cdot I + \B$, where $\L$ and $\B$ are traceless. Then, eqs.~(3.11) and (3.15) are equivalent to, respectively,

$$
B_{-1} = B_{-1t}\cdot I + b_-(t)\L_2,   \eqno(3.16) 
$$

$$
B_1 = B_{1t}\cdot I + b_+(t)\L_0,   \eqno(3.17) 
$$

\ni where $b_+(t)$ and $b_-(t)$ are still arbitrary functions. The trace parts of the other matrix Lax equations give

$$
\prt_tL_{2t} = -B_{-1t},   \eqno(3.18) 
$$

$$
L_{1t} = C_{1t} = \text{const.},   \eqno(3.19) 
$$

$$
\prt_tL_{0t} = B_{1t}.   \eqno(3.20) 
$$

\ni Thus, eqs.~(3.16), (3.18) and (3.17), (3.20) lead, respectively,  to the expressions

$$
B_{-1} = -\prt_tL_{2t}\cdot I + b_-(t)\L_2,   \eqno(3.21) 
$$

$$
B_1 = \prt_tL_{0t}\cdot I + b_+(t)\L_0.   \eqno(3.22) 
$$

\ni The rest becomes the system of three matrix ODEs for traceless matrices:

$$
\prt_t\L_2 = -b_-\L_2 + [\L_2, b_-\L_1- \B_0],   \eqno(3.23) 
$$

$$
\prt_t\L_1 = (b_- - b_+)[\L_2, \L_0] + [\B_0, \L_1],   \eqno(3.24) 
$$

$$
\prt_t\L_0 = b_+\L_0 + [\B_0 - b_+\L_1, \L_0].   \eqno(3.25) 
$$

\ni Now we return to the consequences of FP equation (3.6), using also eqs.~(3.21) and (3.22). One has five matrix equations for five powers of $x$ present in $\D_H(L, B)$:

$$
x^{-2}: L_2^2 - L_2/t = d_{-2}(I-\sigma_3) + a_{-2}\sigma_{-},   \eqno(3.26) 
$$

$$
x^{-1}: \kappa t(b_-\L_2-\prt_tL_{2t}\cdot I) + \{L_2, L_1\} + (a-2)L_2 - L_1/t = d_{-1}(I-\sigma_3) + a_{-1}\sigma_{-},   \eqno(3.27) 
$$

$$
x^{0}: \kappa tB_0 + L_1^2 + \{L_2, L_0\} + (a-1)L_1 - L_2 - L_0/t = d_{0}(I-\sigma_3) + a_{0}\sigma_{-},   \eqno(3.28) 
$$

$$
x^{1}: \kappa t(\prt_tL_{0t}\cdot I + b_+(t)\L_0) + \{L_1, L_0\} + aL_0 - L_1 = d_{1}(I-\sigma_3) + a_{1}\sigma_{-},   \eqno(3.29) 
$$

$$
x^{2}: L_0^2 - L_0 = d_{2}(I-\sigma_3) + a_{2}\sigma_{-}.   \eqno(3.30) 
$$

\ni The outcome of the last equations is 10 scalar relations (the rest just determine the 10 arbitrary functions on the r.h.s.). Among them there are 6 algebraic constraints on the components of $L$-matrices:

$$
(L_{2t} + L_{2d})^2 - (L_{2t} + L_{2d})/t + L_{2a}^2 = 0,   \eqno(3.31) 
$$

$$
(2L_{2t} - 1/t)L_{2+} = 0,   \eqno(3.32) 
$$

$$
(L_{0t} + L_{0d})^2 - (L_{0t} + L_{0d}) + L_{0a}^2 = 0,   \eqno(3.33) 
$$

$$
(2L_{0t} - 1)L_{0+} = 0,   \eqno(3.34) 
$$

$$
(\kappa tb_- + 2C_{1t} + a - 2)L_{2+} + (2L_{2t} - 1/t)L_{1+} = 0,   \eqno(3.35) 
$$

$$
(\kappa tb_+ + 2C_{1t} + a)L_{0+} + (2L_{0t} - 1)L_{1+} = 0,   \eqno(3.36) 
$$

\ni where we decomposed the traceless matrices as $\L = L_d\sigma_3 + L_a = L_d\sigma_3 + L_+\sigma_+ + L_-\sigma_-$. The other equations are 2 ODEs for the traces:

$$
\kappa t\prt_tL_{2t} = (\kappa tb_- + 2C_{1t} + a - 2)L_{2d} + (a-2)L_{2t} + (C_{1t} + L_{1d})(2L_{2t} - 1/t) + \{\L_2, \L_1\},   \eqno(3.37) 
$$

$$
\kappa t\prt_tL_{0t} = -(\kappa tb_+ + 2C_{1t} + a)L_{0d} - aL_{0t} - (C_{1t} + L_{1d})(2L_{0t} - 1) - \{\L_1, \L_0\},   \eqno(3.38) 
$$

\ni and two expressions for two components of the matrix $B_0$:

$$
B_+ \equiv \kappa tB_{0+} = (1-2L_{0t})L_{2+} - (2C_{1t} + a - 1)L_{1+} + (1/t - 2L_{2t})L_{0+},   \eqno(3.39) 
$$

$$
B_t \equiv \kappa tB_{0t} = -\L_1^2 - \{\L_2, \L_0\} - 2L_{2t}L_{0t} + L_{2t} + L_{0t}/t - C_{1t}(C_{1t}+a-1) + d_0,   \eqno(3.40) 
$$

\ni eq.~(3.39) being another constraint, while eq.~(3.40) just determines $B_{0t}$ in terms of $B_{0d}$ since it follows from eq.~(3.28) that

$$
\hat B \equiv \kappa t\B_0 = B_+\sigma_+ + B_d\sigma_3 + B_-\sigma_- = (1-2L_{0t})\L_2 - (2C_{1t} + a - 1)\L_1 + (1/t - 2L_{2t})\L_0 - d_0\sigma_3 + a_0\sigma_-.   \eqno(3.41) 
$$

\ni Since $B_{0t}$ does not enter any other equations, eq.~(3.40) has no other consequences. As for the Lax eqs.~(3.23)--(3.25), for the following it is convenient to rewrite them in terms of just introduced matrix $\hat B$, whose component $B_+$ is determined by eq.~(3.39) and the other two components $B_d$ and $B_-$ are still arbitrary:

$$
\kappa t\prt_t\L_2 = -\kappa tb_-\L_2 + [\L_2, \kappa tb_-\L_1- \hat B],   \eqno(3.42) 
$$

$$
\kappa t\prt_t\L_1 = \kappa t(b_- - b_+)[\L_2, \L_0] + [\hat B, \L_1],   \eqno(3.43) 
$$

$$
\kappa t\prt_t\L_0 = \kappa tb_+\L_0 + [\hat B - \kappa tb_+\L_1, \L_0].   \eqno(3.44) 
$$

\ni Thus, it is worth mentioning that, as a result of the above considerations, one is left with 4 arbitrary functions of $t$: $b_-$, $b_+$, $B_d$ and $B_-$ entering the further analysis. The rest essentially depends on the way of resolving the constraints -- eqs.~(3.31)--(3.36).

\section{Different cases of constraint resolution}

\par {\bf\large Case I:} Resolve eqs.~(3.32), (3.34) as

$$
L_{2t} = \frac{1}{2t}, \ \ \ \ \ L_{0t} = \frac{1}{2},   \eqno(4.1) 
$$

\ni Then in place of eqs.~(3.31) and (3.33) one gets, respectively, 

$$
\L_2^2 = L_{2d}^2 + L_{2a}^2 = \frac{1}{4t^2},   \eqno(4.2) 
$$

 $$
\L_0^2 = L_{0d}^2 + L_{0a}^2 = \frac{1}{4}.   \eqno(4.3) 
$$

\ni One then compares these with the consequences of anti-commutation of $\L_2$ with eq.~(3.42) and $\L_0$ with eq.~(3.44), respectively:

$$
\kappa t\prt_t\L_2^2 = -2\kappa tb_-\L_2^2,   \eqno(4.4) 
$$

$$
\kappa t\prt_t\L_0^2 = 2\kappa tb_+\L_0^2,   \eqno(4.5) 
$$

\ni from where it follows that two of the four arbitrary functions are not so anymore:

$$
tb_- = 1, \ \ \ \ \ b_+=0.   \eqno(4.6) 
$$

\ni Then the constraints (3.35) and (3.36) read:

$$
(\kappa - 2 + 2C_{1t} + a)L_{2+} = 0,  \eqno(4.7) 
$$ 
  
$$
(2C_{1t} + a)L_{0+} = 0.  \eqno(4.8) 
$$

\ni Moreover, now the trace eqs.~(3.37) and (3.38) become algebraic:

$$
(\kappa - 2 + 2C_{1t} + a)L_{2d} = -\{\L_2, \L_1\} - \frac{\kappa-2+a}{2t},  \eqno(4.9) 
$$

$$
(2C_{1t} + a)L_{0d} = -\{\L_1, \L_0\} - \frac{a}{2}.  \eqno(4.10) 
$$

\ni The eq.~(3.41) now becomes

$$
\hat B = -(2C_{1t} + a - 1)\L_1 + B_d\sigma_3 + B_-\sigma_-.  \eqno(4.11) 
$$

\par {\bf\large Case II:} Resolve eqs.~(3.32), (3.34) as

$$
L_{2t} = \frac{1}{2t}, \ \ \ \ \ L_{0+} = 0,   \eqno(4.12) 
$$

\ni considering $L_{0t} \neq 1/2$. Then eqs.~(3.31) and (3.33) turn, respectively, into eq.~(4.2) and

$$
(L_{0t}+L_{0d})(L_{0t}+L_{0d}-1) = 0,   \eqno(4.13)
$$

\ni eq.~(3.35) becomes

$$
(\kappa tb_- - 2 + 2C_{1t} + a)L_{2+} = 0,  \eqno(4.14) 
$$ 

\ni and eq.~(3.36) gives

$$
L_{1+} = 0.   \eqno(4.15)
$$

\ni Taking also $L_{2+}=0$ would give a subcase of Case IV below, so we choose $\kappa tb_- - 2 + 2C_{1t} + a = 0$ in eq.~(4.14). Eq.~(3.37) becomes again algebraic,

$$
\{\L_2, \L_1\} = -\frac{\kappa-2+a}{2t},  \eqno(4.16) 
$$

\ni while eq.~(3.38) remains intact so far. Eq.~(3.39) simplifies to 

$$
B_+ \equiv \kappa tB_{0+} = (1-2L_{0t})L_{2+},   \eqno(4.17) 
$$

\ni with right-hand side not identically zero. 








Then it is easy to verify that under the above conditions ODEs (3.43) and (3.44) can be solved only by taking $\L_1=0$ and $L_0=0$. Now the relation $\kappa-2+a=0$ follows from eq.~(4.16), and this is the equation (2.1) if one returns to the original $a$ (recall that we redenoted $a+2/\beta= a+1/\kappa \to a$ before).
\par Eq.~(4.2) and anticommutator of eq.~(3.42) with $\L_2$ imply $tb_-=1$ again and then $C_{1t}=0$ follows from eqs.~(4.14) and (4.16). 
Eq.~(4.13) and $\L_0=0$ give $L_{0t}(L_{0t}-1)=0$, and this together with other above facts entails $L_{0t}=0$ as solution of eq.~(3.38). Thus, only eq.~(4.2) and eq.~(3.42) together with eq.~(4.17), $B_+=L_{2+}$, remain, containing two arbitrary functions, which ultimately leads to the Lax pair from Theorem 3,  which makes sense for $a \le 0$ and {\it any} (nonnegative) value of $\beta$. 

\par {\bf\large Case III:} Resolve eqs.~(3.32), (3.34) as

$$
L_{2+} = 0, \ \ \ \ \ L_{0t} = \frac{1}{2}.   \eqno(4.24) 
$$

\ni It turns out that Case III leads to a special type solution for which $\kappa =2$ and $a=0$, which can be seen similarly to the above consideraton of Case II. This seems not very interesting since we have more general solution for $\kappa=2$ and arbitrary admissible $a$. 



{\bf\large Case IV:} Resolve eqs.~(3.32), (3.34), (3.35) and (3.36) as

$$
L_{2+} = L_{1+} = L_{0+} = 0,   \eqno(4.25)   
$$

\ni Then eq.~(3.39) gives $B_+=0$, while eqs.~(3.31) and (3.33) become, respectively,

$$
(L_{2t}+L_{2d})(L_{2t}+L_{2d}-1/t) = 0,   \eqno(4.26)   
$$

$$
(L_{0t}+L_{0d})(L_{0t}+L_{0d}-1) = 0.   \eqno(4.27)  
$$
 
\ni The `+'-components of eqs.~(3.42)-(3.44) become now trivial, while their diagonal components reduce to

$$
\kappa tL_{2d}' = -\kappa tb_-(t)L_{2d},  \eqno(4.28)  
$$

$$
L_{1d} = C_{1d} = \text{ const.},   \eqno(4.29)  
$$

$$
\kappa tL_{0d}' = \kappa tb_+(t)L_{0d},  \eqno(4.30)  
$$

\ni respectively. Adding trace ODE eq.~(3.35) with eq.~(4.28), taking into account eqs.~(4.25) and (4.29), gives

$$
\kappa t(L_{2t}+L_{2d})' = (2(C_{1t}+C_{1d}) + a -2)(L_{2t}+L_{2d}) - (C_{1t}+C_{1d})/t,   \eqno(4.31)  
$$

\ni and adding trace ODE eq.~(3.36) with eq.~(4.30), taking into account eqs.~(4.25) and (4.29), gives

$$
\kappa t(L_{0t}+L_{0d})' = -(2(C_{1t}+C_{1d}) + a)(L_{0t}+L_{0d}) + (C_{1t}+C_{1d}).   \eqno(4.32)  
$$

\ni However, eq.~(4.27) implies that $(L_{0t}+L_{0d})'=0$, and eq.~(4.32) becomes

$$
(2(C_{1t}+C_{1d}) + a)(L_{0t}+L_{0d}) = C_{1t}+C_{1d}.   \eqno(4.33)  
$$

\ni Further consideration of the above constraints leads one to the conclusion that in this case there are only possibilities similar to Cases II and III above.

\section{Painlev\'e equations for $\beta= 2$ and $4$}  

It is convenient to introduce

$$
X_2 = t\L_2.    \eqno(5.1)
$$

\subsection{Case Ia} 

Suppose $2C_{1t} + a \neq 0$. Then eq.~(4.8) gives $L_{0+} = 0$, therefore eq.~(4.3) becomes $L_{0d}^2 = \frac{1}{4}$. So let

$$
2L_{0d} = C_{0d}, \ \ \ \ \ C_{0d} = \pm 1.   \eqno(5.2)  
$$

\ni The ``+"-component of eq.~(3.44) now reads $(2C_{1t} + a - 1)L_{1+} = 0$. If $L_{1+} = 0$, then the ``+"-component of eq.~(3.43) gives $L_{2+}=0$, and this reduces to a subcase of Case IV. So we take

$$
2C_{1t} + a - 1 = 0.  \eqno(5.3) 
$$

\ni Then eq.~(4.7) becomes $(\kappa-1)L_{2+}=0$, and again taking $L_{2+} = 0$ reduces to a subcase of Case IV since then the ``+"-component of eq.~(3.42) gives also $L_{1+}=0$. Thus, we get $\kappa = 1$ (i.e.~$\beta=2$). We introduce new variables:



$$
L_{1+} = \phi,  \ \ \ y = \frac{\phi'}{\phi} - 2\frac{B_d}{t}, \ \ \ l_- = \phi L_{0-}, \ \ \ A_- = \phi B_-, \ \ \ A_1= L_{1a}^2=\phi L_{1-}.   \eqno(5.4)  
$$







\ni The system of ODEs (3.42)-(3.44) is now

$$
t\prt_tX_2 = [X_2, \L_1-\hat B],   \eqno(5.5)  
$$

$$
t^2\prt_t\L_1 = [X_2, \L_0]+t[\hat B, \L_1],   \eqno(5.6)  
$$

$$
t\prt_t\L_0 = [\hat B, \L_0].   \eqno(5.7)  
$$






 

\ni The diagonal component of eq.~(5.7) is redundant, and its `+'-component is trivial. The only nontrivial component of eq.~(5.7) reads 




$$
t\prt_tl_- = t\frac{\phi'}{\phi}l_- + C_{0d}A_- - 2B_dl_-.   \eqno(5.8)  
$$



\ni The components of eqs.~(5.5), (5.6) become 

$$
t\prt_tX_{2d} = \frac{X_{2+}}{\phi}(A_1-A_-) - \phi X_{2-},   \eqno(5.5d)  
$$

$$
t\prt_tX_{2+} = 2\phi X_{2d} - 2(L_{1d}-B_d)X_{2+},   \eqno(5.5+)  
$$

$$
t\prt_tX_{2-} =  2(L_{1d}-B_d)X_{2-} - 2\frac{X_{2d}}{\phi}(A_1-A_-),   \eqno(5.5-)  
$$

$$
t^2\prt_tL_{1d} = \frac{X_{2+}}{\phi}l_- - tA_-,   \eqno(5.6d)  
$$

$$
t^2\prt_t\phi = -C_{0d}X_{2+} + 2t\phi B_d,   \eqno(5.6+)  
$$

$$
t^2\prt_tA_1 = t^2\frac{\phi'}{\phi}A_1 + C_{0d}\phi X_{2-} - 2X_{2d}l_- - 2t(B_dA_1 - L_{1d}A_-).   \eqno(5.6-)  
$$








\ni We now eliminate $X_{2+}$, expressing it from eq.~(5.6+) and using $y$ from eq.~(5.4),

$$
X_{2+} = -C_{0d}t^2y\phi.   \eqno(5.9)  
$$

\ni Substituting it into the constraint eqs.~(4.2), (4.9) gives, respectively,

$$
4(X_{2d}^2 - C_{0d}t^2y\phi X_{2-})=1,   \eqno(5.10)  
$$

$$
2L_{1d}X_{2d} - C_{0d}t^2yA_1 + \phi X_{2-} = -\frac{a-1}{2}.  \eqno(5.11)  
$$

\ni The remaining constraint to take into account, eq.~(4.10), reduces to

$$
C_{0d}L_{1d} + l_- = - \frac{a+C_{0d}}{2}.  \eqno(5.12)  
$$

\ni The ODE (5.6d) becomes, after substituting eq.~(5.9) and using $y$-variable,

$$
t\prt_tL_{1d} = -C_{0d}tyl_- - A_-.  \eqno(5.13)  
$$






\ni In fact the algebraic equations (5.10), (5.11) and (5.12) turn out to be first integrals of the remaining ODEs. Therefore it is possible and convenient to consider eqs.~(5.5-) and (5.6-) as redundant. We next express $l_-$ from eq.~(5.12)  and substitute it and $y$-variable into eq.~(5.8), which turns into

$$
t\prt_tL_{1d} = ty(L_{1d}+(C_{0d}a+1)/2) - A_-,   \eqno(5.14)  
$$

\ni which is identical to eq.~(5.13) after eliminating $l_-$ from the last by eq.~(5.12). We express $X_{2-}$ from eq.~(5.10), $X_{2-} = \frac{4X_{2d}^2-1}{4C_{0d}t^2y\phi}$, plug it into eq.~(5.11) and express $A_1$ from there,


 
$$
C_{0d}t^2yA_1 = \frac{4X_{2d}^2-1}{4C_{0d}t^2y} + 2L_{1d}X_{2d} + \frac{a-1}{2}.  \eqno(5.15)  
$$

\ni After introducing

$$
r = -2C_{0d}X_{2d}, \ \ \ \ \ l_d = L_{1d} + (C_{0d}a+1)/2,   \eqno(5.16)  
$$

\ni we substitute expressions for $X_{2-}$ and $A_1$ into eq.~(5.5d), and the last becomes 


$$
tr' = -2t^2yA_- + \frac{r^2-1}{t^2y} - (2l_d-C_{0d}a-1)r + C_{0d}(a-1).   \eqno(5.17)  
$$





\ni With the substitutions from eqs.~(5.4), (5.9) and (5.16) done, eq.~(5.5+) can be represented as

$$
t(t^2y)' = -t^3y^2 + r - (2l_d-C_{0d}a-1)t^2y,   \eqno(5.18)  
$$



\ni and eq.~(5.14) now reads

$$
tl_d' = tyl_d - A_-.   \eqno(5.19)  
$$

\ni Thus, we are left with eqs.~(5.17), (5.18) and (5.19). There is stil one arbitrary function, $A_-$, left ($B_d$ was already absorbed), but this can be effectively absorbed too. We differentiate eq.~(5.18) to get

$$
t(t^2y)'' = - 2t^3yy' - 3t^2y^2 + r' - 2t^2yl_d' - (2l_d-C_{0d}a)(t^2y)'.   \eqno(5.20)  
$$

\ni Then we subtract eq.~(5.19) multiplied by $2t^2y$ from eq.~(5.17), which cancels $A_-$ there, and obtain

$$
t(r' - 2t^2yl_d') = -2t^3y^2 + \frac{r^2-1}{t^2y} - (2l_d-C_{0d}a-1)r + C_{0d}(a-1).   \eqno(5.21)  
$$

\ni Now we plug the right-hand sides of eq.~(5.21) into eq.~(5.20) multiplied by $t$:

$$
t^2(t^2y)'' = - 2t^4yy' - 3t^3y^2 - 2t^3y^2 + \frac{r^2-1}{t^2y} - (2l_d-C_{0d}a-1)r + C_{0d}(a-1) - (2l_d-C_{0d}a)t(t^2y)'.   \eqno(5.22)  
$$

\ni Next we put the expression for $r$ from eq.~(5.18) in eq.~(5.22). Then all terms containing $l_d$ cancel and we obtain equation for $y$:

$$
y'' = \frac{(y')^2}{y} - \frac{1}{t}y' + y^3 - \frac{C_{0d}a}{t}y^2 + \frac{C_{0d}(a-1)}{t^4} - \frac{1}{t^6y}.   \eqno(5.23)  
$$

\ni This last equation is in fact a form of Painlev\'e III. To bring it to the standard form of Painlev\'e III, see e.g.~\cite{JM81, F2010}, one just needs to define

$$
\xi = t^{-1/2}, \ \ \ \ \ y = t^{-3/2}z = \xi^3z,
$$

\ni then, in the new variables, eq.~(5.23) is a standard Painlev\'e III~\cite{JM81, F2010}:

$$
z''(\xi) = \frac{(z'(\xi))^2}{z} - \frac{1}{\xi}z'(\xi) + 4z^3 - \frac{4C_{0d}a}{\xi}z^2 + \frac{4C_{0d}(a-1)}{\xi} - \frac{4}{z}.   \eqno(5.24)  
$$

\ni If one puts $C_{0d} = -1$, then eq.~(5.23) becomes exactly eq.~(2.7a) from Theorem 1. 


\subsection{Case Ib} 

Now take $2C_{1t} + a = 0$. Then eq.~(4.8) is trivial. Using eq.~(5.1), one obtains $(\kappa - 2)X_{2+} = 0$ from eq.~(4.7). Eq.~(3.41) now becomes $\hat B = L_{1+}\sigma_+ + B_d\sigma_3 + B_-\sigma_-$, where functions $B_d(t)$ and $B_-(t)$ are arbitrary. If $\kappa \neq 2$, then $X_{2+} = 0$ here. Therefore eq.~(4.2) becomes $X_2^2 = X_{2d}^2 = \frac{1}{4}$, while the `+'-component of eq.~(3.42) reads simply $(\kappa-1)X_{2d}L_{1+} = 0$, which gives again $\kappa = 1$, since taking $L_{1+} = 0$ leads again to a subcase of Case IV. The further consideration is similar to the Case I-I. Now the system of ODEs (3.42)-(3.44) is

  







$$
t\prt_tX_2 = [X_2, \L_1 - \hat B],   \eqno(5.25)  
$$

$$
t^2\prt_t\L_1 = [X_2, \L_0] + t[\hat B, \L_1],   \eqno(5.26)  
$$

$$
t\prt_t\L_0 = [\hat B, \L_0].   \eqno(5.27)  
$$

\ni We again introduce new variables:

$$
L_{1+} = \phi,  \ \ \ y = \frac{\phi'}{\phi} - 2\frac{(B_d-L_{1d})}{t}, \ \ \ l_- = \phi X_{2-}, \ \ \ A_- = \phi B_-, \ \ \ A_1= L_{1a}^2=\phi L_{1-}.   \eqno(5.28)  
$$

\ni Let also 

$$
2X_{2d} = C_{2d}, \ \ \ \ \ C_{2d} = \pm 1,   \eqno(5.29)  
$$

\ni then the only nontrivial `-'-component of eq.~(5.25) reads


$$
tl_-' = t\frac{\phi'}{\phi}l_- + 2(L_{1d}-B_d)l_- - C_{2d}(A_1-A_-).   \eqno(5.30)  
$$

\ni The components of eqs.~(5.26), (5.27) become, respectively, 







$$
t^2\prt_tL_{1d} = -\frac{L_{0+}}{\phi}l_- + t(A_1-A_-),   \eqno(5.26d)  
$$

$$
t^2\prt_t\phi = C_{2d}L_{0+} + 2t\phi (B_d-L_{1d}),   \eqno(5.26+)  
$$

$$
t^2\prt_tA_1 = t^2\frac{\phi'}{\phi}A_1 + 2L_{0d}l_- - C_{2d}\phi L_{0-} - 2t(B_dA_1 - L_{1d}A_-),   \eqno(5.26-)  
$$

$$
t\prt_tL_{0d} = \phi L_{0-} - \frac{L_{0+}}{\phi}A_-,   \eqno(5.27d)  
$$

$$
t\prt_tL_{0+} = 2B_dL_{0+} - 2\phi L_{0d},   \eqno(5.27+)  
$$

$$
t\prt_tL_{0-} =  2\frac{L_{0d}}{\phi}A_- - 2B_dL_{0-} ,   \eqno(5.27-)  
$$


\ni Three among six components of eqs.~(5.26) and (5.27) are again redundant due to constraints-first integrals (4.3), (4.9) and (4.10). The algebraic equations (5.32), (5.33) and (5.34) below are first integrals of the remaining ODEs. Therefore it is possible and convenient to consider eqs.~(5.26-) and (5.27-) as redundant. We eliminate $L_{0+}$, expressing it from eq.~(5.26+) and using $y$ from eq.~(5.28), 


 



$$
L_{0+} = C_{2d}t^2y\phi.   \eqno(5.31)  
$$

\ni Substituting it (and notations from eq.~(5.28)) into the constraint eqs.~(4.3), (4.10) gives, respectively,

$$
4(L_{0d}^2 + C_{2d}t^2y\phi L_{0-})=1,   \eqno(5.32)  
$$

$$
2L_{1d}L_{0d} + \phi L_{0-} + C_{2d}t^2yA_1 = -\frac{a}{2}.  \eqno(5.33)  
$$

\ni The remaining constraint to consider, eq.~(4.9) reads now

$$
C_{2d}L_{1d} + l_- = - \frac{a-1-C_{2d}}{2},  \eqno(5.34)  
$$

\ni The ODE (5.26d) becomes, after substituting eq.~(5.31) and using $y$-variable,

$$
t\prt_tL_{1d} = -C_{2d}tyl_- + A_1-A_-.   \eqno(5.35)  
$$







\ni We next express $l_-$ from eq.~(5.34)  and substitute it and variable $y$ from eq.~(5.4) into eq.~(5.30), which turns into

$$
t\prt_tL_{1d} = ty\left(L_{1d} + C_{2d}\cdot\frac{a-1-C_{2d}}{2}\right) + A_1-A_-,   \eqno(5.36)  
$$

\ni which is identical to eq.~(5.35) after eliminating $l_-$ from the last by eq.~(5.34). We express $L_{0-}$ from eq.~(5.32), $L_{0-} = -\frac{4L_{0d}^2-1}{4C_{2d}t^2y\phi}$, then plug it into eq.~(5.33) and express $A_1$ from there,


 
$$
t^2yA_1 = \frac{4L_{0d}^2-1}{4t^2y} - 2C_{2d}L_{1d}L_{0d} - \frac{C_{2d}a}{2}.  \eqno(5.37)  
$$

\ni Now we make the last change of variables, introduce

$$
r = -2C_{2d}L_{0d}, \ \ \ \ \ l_d = L_{1d} + (C_{2d}(a-1) - 1)/2.   \eqno(5.38)  
$$



\ni With the substitutions from eqs.~(5.28), (5.31) and (5.38) performed,  eq.~(5.27+) transforms into

$$
t(t^2y)' = -t^3y^2 + t^2y(2l_d - C_{2d}a + C_{2d} + 1) + r,   \eqno(5.39)  
$$

\ni When we substitute the expressions (5.31), (5.32) and (5.38) into eq.~(5.27d), the last becomes 

$$
tr' = \frac{r^2-1}{2t^2y} + 2t^2yA_-.   \eqno(5.40)  
$$

\ni Next we plug expression (5.37) for $A_1$ into eq.~(5.36) and use (5.38) there to obtain   

$$
t^3yl_d' = (t^3y^2+r)l_d + \frac{r^2-1}{4t^2y} - \frac{C_{2d}a}{2}(r+1) + \frac{(C_{2d}+1)}{2}r - t^2yA_-.   \eqno(5.41)  
$$

\ni Thus, we are left with eqs.~(5.39), (5.40) and (5.41). There is still one arbitrary function, $A_-$, left ($B_d$ was already absorbed), but this can be absorbed too. We differentiate eq.~(5.39) to get

$$
t(t^2y)'' = - 2t^3yy' - 3t^2y^2 + r' + 2t^2yl_d' + (t^2y)'(2l_d-C_{2d}a+C_{2d}).   \eqno(5.42)  
$$

\ni Then we add eq.~(5.40) and eq.~(5.41) multiplied by $2$, which cancels terms with $A_-$ there, and obtain

$$
t(r' + 2t^2yl_d') = \frac{r^2-1}{t^2y} + 2(t^3y^2 + r)l_d - C_{2d}a(r+1) + (C_{2d}+1)r.   \eqno(5.43)  
$$

\ni Now we plug the right-hand sides of eq.~(5.43) into eq.~(5.42) multiplied by $t$:

$$
t^2(t^2y)'' = - 2t^4yy' - 3t^3y^2 + \frac{r^2-1}{t^2y} + 2(t^3y^2 + r)l_d - C_{2d}a(r+1) + (C_{2d}+1)r + t(t^2y)'(2l_d-C_{2d}a+C_{2d}).   \eqno(5.44) 
$$

\ni We then express $l_d$ from eq.~(5.39) and put it into eq.~(5.44). All the terms containing $r$ then happily cancel and an ODE for $y$ results:  



$$
y'' = \frac{(y')^2}{y} - \frac{1}{t}y' + y^3 + \frac{C_{2d}(a-1)}{t}y^2 - \frac{C_{2d}a}{t^4} - \frac{1}{t^6y}.   \eqno(5.45)  
$$

\ni This last equation is in fact a form of Painlev\'e III. To bring it to the standard form of Painlev\'e III, see e.g.~\cite{JM81, F2010}, one just needs to define

$$
\xi = t^{-1/2}, \ \ \ \ \ y = t^{-3/2}z = \xi^3z,
$$

\ni then, in the new variables, eq.~(5.45) is a standard Painlev\'e III~\cite{JM81, F2010}:

$$
z''(\xi) = \frac{(z'(\xi))^2}{z} - \frac{1}{\xi}z'(\xi) + 4z^3 + \frac{4C_{2d}(a-1)}{\xi}z^2 - \frac{4}{z} - \frac{4C_{2d}a}{\xi}.   \eqno(5.46)  
$$

\ni If one puts $C_{2d} = -1$, then eq.~(5.45) becomes exactly eq.~(2.7b) from Theorem 1. 

\subsection{Case Ic} 

Also take $2C_{1t} + a = 0$ but resolve eq.~(3.35) as $\kappa = 2$, then eqs.~(3.37) and (3.38) become, respectively,


  
$$
\{X_2, \L_1\} = - \frac{a}{2},  \eqno(5.47)  
$$

$$
\{\L_1, \L_0\} = - \frac{a}{2}.  \eqno(5.48)  
$$

\ni Take now the two arbitray functions in eq.~(3.41) to be $d_0=a_0=0$. Then eq.~(3.41) simplifies to $\hat B = \L_1$. The triple of ODEs (3.42)--(3.44) now reads, when also taking into account eq.~(5.1),



$$
2t\prt_tX_2 = [X_2, \L_1], \ \ \ \ \ t^2\prt_t\L_1 = [X_2, \L_0], \ \ \ \ \ 2t\prt_t\L_0 = [\L_1, \L_0].  \eqno(5.49)  
$$



\ni Let us now change variables since we want to use some symmetry between $X_2$ and $\L_0$, i.e.~let

$$
S = X_2 + \L_0, \ \ \ \ \ A = X_2 - \L_0.   \eqno(5.50)  
$$

\ni One can rewrite all the remaining equations in terms of $\L_1$, $S$, and $A$. Recall also constraint eqs.~(4.2) and (4.3). Their sum and difference give, respectively,

$$
S^2 + A^2 = 1,  \ \ \ \ \ \{S, A\} = 0,  \eqno(5.51)  
$$


\ni while the sum and difference of eqs.~(5.47) and (5.48) give

$$
\{S, \L_1\} = -a, \ \ \ \ \ \{\L_1, A\} = 0.   \eqno(5.52)  
$$


\ni At last, the system of ODEs (5.49) is equivalent to the following system:  

$$
2t\prt_tS = [A, \L_1],   \eqno(5.53)  
$$

$$
2t^2\prt_t\L_1 = -[S, A],   \eqno(5.54)  
$$

$$
2t\prt_tA = [S, \L_1].   \eqno(5.55)  
$$

\ni Next we rewrite the system in diagonal-anti-diagonal form ($S = S_d\sigma_3 + S_a$ etc.):

$$
S_d^2 + S_a^2 + A_d^2 + A_a^2 = 1, \ \ \ \ \ 2S_dA_d + \{S_a, A_a\} = 0,  \eqno(5.51)  
$$


$$
2S_dL_{1d} + \{S_a, L_{1a}\} = -a, \ \ \ \ \ 2L_{1d}A_d + \{L_{1a}, A_a\} = 0,  \eqno(5.52)  
$$


$$
2t\prt_tS_d = [A_a, L_{1a}]\sigma_3,   \eqno(5.53d)  
$$

$$
t\prt_tS_a = A_d\sigma_3L_{1a} - L_{1d}\sigma_3A_a,   \eqno(5.53a)  
$$

$$
2t^2\prt_tL_{1d} = -[S_a, A_a]\sigma_3,   \eqno(5.54d)  
$$

$$
t^2\prt_tL_{1a} = A_d\sigma_3S_a - S_d\sigma_3A_a,   \eqno(5.54a)  
$$

$$
2t\prt_tA_d = [S_a, L_{1a}],   \eqno(5.55d)  
$$

$$
t\prt_tA_a = S_d\sigma_3L_{1a} - L_{1d}\sigma_3S_a   \eqno(5.55a)  
$$

\ni It turns out that likely a simplest way to find the solution here is to put $A_a = 0$. Then the system is greatly simplified and its non-trivial solution can be found by choosing also

$$
S_d = L_{1d} = 0.
$$

\ni Then the second equations in eqs.~(5.51), (5.52), as well as eqs.~(5.53d), (5.54d) and (5.55a) are trivially satisfied and what remains is

$$
S_a^2 + A_d^2 = 1,   \eqno(5.56)  
$$

$$
\{S_a, L_{1a}\} = -a,   \eqno(5.57)  
$$

$$
t\prt_tS_a = A_d\sigma_3L_{1a},   \eqno(5.58)  
$$

$$
t^2\prt_tL_{1a} = A_d\sigma_3S_a,   \eqno(5.59)  
$$

$$
2t\prt_tA_d = [S_a, L_{1a}].   \eqno(5.60)  
$$

\ni Now, after expressing $S_a$ from eq.~(5.59) and plugging it into eqs.~(5.56), (5.57) and (5.60), they turn into, respectively,

$$
t^4(\prt_tL_{1a})^2 = A_d^2(A_d^2 - 1),   \eqno(5.61)  
$$

$$
t^2[L_{1a}, \prt_tL_{1a}]\sigma_3 = aA_d,   \eqno(5.62)  
$$

$$
\prt_t(A_d^2) = t\prt_t(L_{1a}^2).   \eqno(5.63)  
$$

\ni Differentiating e.g.~eq.~(5.57) and using eq.~(5.58) one can see that eq.~(5.58) can be considered as redundant -- it is a consequence of the others. To close the system of the last three written out equations, it is convenient to use the identity:

$$
\{L_{1a}, \prt_tL_{1a}\}^2 - [L_{1a}, \prt_tL_{1a}]^2 = 4L_{1a}^2(\prt_tL_{1a})^2.   \eqno(5.64)  
$$

\ni We substitute eqs.~(5.61), (5.63) and the square of eq.~(5.62) into eq.~(5.64) and obtain:

$$
4(A_d^2-1)L_{1a}^2 = 4(t\prt_tA_d)^2 - a^2.   \eqno(5.65)  
$$

\ni Next we differentiate eq.~(5.65) and find:

$$
4(A_d^2-1)\prt_tL_{1a}^2 + 8A_d\prt_tA_dL_{1a}^2 = 8(t\prt_tA_d)\prt_t(t\prt_tA_d).   \eqno(5.66)  
$$

\ni Finally we substitute $L_{1a}^2$ from eq.~(5.65) and $\prt_tL_{1a}^2$ from eq.~(5.63) into eq.~(5.66), which yields an ODE for the single variable $A_d$:

$$
t(A_d^2-1)(tA_d')' = A_d(tA_d')^2 + \frac{1}{t}A_d^3(A_d^2-2) + \left(\frac{1}{t} - \frac{a^2}{4}\right)A_d,   \eqno(5.67)  
$$

\ni where again $' \equiv \prt_t$. One can now recognize the last equation as almost identical to the equation (1.16) of the paper~\cite{TW-Bes} about the hard edge of unitary-invariant random matrix ensembles (there $\beta = 2$) and Bessel kernel. By changing the independent variable in eq.~(5.67) to $\xi = 4/t$, the last becomes exactly that equation from~\cite{TW-Bes}, which is shown there to be a simple transformation of a sigma-form of Painlev\'e III: setting $A_d^2/4 = d(\xi R(\xi))/d\xi$, one can show that $R(\xi)$ satisfies a sigma-form of Painlev\'e III~\cite{TW-Bes}, i.e.~this $\xi R(\xi)$ is a particular classical Painlev\'e III Hamiltonian.
\par Thus, for $\beta=4$ hard edge we also have a Painlev\'e III representation.

\section{Explicit Lax Pairs}

\subsection{Case Ia} 

Here $\beta=2$, $L_{0+}=0$, $L_{0d}^2=1/4$ and $C_{1t} = (1-a)/2$. Anticipating a simpler result, we put $B_d=B_a=0$ and $L_{0d}=-1/2$. From the `-'-component of eq.~(5.7) it follows that $L_{0-} = \text{const}$ and we put $L_{0-}=0$, in order to get the simplest solution. 
Then we have $X_{2d}=r/2$ from eq.~(5.16), and eq.~(5.12) gives 

$$
L_{1d} = \frac{a-1}{2},   \eqno(6.1)  
$$



\ni Clearly now eq.~(5.6d) is satisfied by eq.~(5.12). It is convenient to express the antidiagonal components $X_{2a}$ from eq.~(5.6), $X_{2a} = t^2\sigma_3L_{1a}'$. Substituting the last expression into eqs.~(4.2), (4.9) and (5.5d) yields, respectively, 








$$
t^4(L_{1a}')^2 = \frac{1}{4}(r^2-1),   \eqno(6.2)  
$$

$$
t^2[L_{1a}, L_{1a}']\sigma_3 = \frac{a-1}{2}(r+1),    \eqno(6.3)  
$$

$$
2t(L_{1a}^2)' = r'.  \eqno(6.4)  
$$

\ni Now we use the identity  eq.~(5.64) and plug the last three equations into it to derive

$$  
4(r^2-1)L_{1a}^2 = (tr')^2 - (a-1)^2(r+1)^2.   \eqno(6.5)   
$$

\ni This is to be used together with eq.~(6.4) to eliminate $L_{1a}^2$ and obtain an ODE for $r$. We differentiate eq.~(6.5) and, comparing with eq.~(6.4), we obtain the equation satisfied by $r$,



$$
t(r^2-1)(tr')' = r(tr')^2 + \frac{1}{t}(r^2-1)^2 - (a-1)^2(r+1)^2.   \eqno(6.6)  
$$

\ni If we introduce function $q$ such that 

$$
r = -1 + 2q^2,   \eqno(6.7)
$$

\ni which is inspired by the hard-to-soft edge limit, see below, then the new $q$ will satisfy

$$
t(q^2-1)(tq')' = q(tq')^2 + \frac{q}{t}(q^2-1)^2 - \frac{(a-1)^2}{4}q,    \eqno(6.8)  
$$

\ni which is nothing but an instance of the Tracy-Widom equation (5.67) (up to change $a-1 \to a$) for $\beta=2$ hard edge. 
\par Now we write out the explicit Lax pair for this case using the last findings and the previous consideration ending in a standard Painlev\'e III for $y = L_{1+}'/L_{1+}$, see eq.~(5.23) (one should put $C_{0d}=-1$ there to be consistent with the current consideration). First we recall that $X_2=t\L_2$ and so $L_{2d} = X_{2d}/t = r/(2t)$ and $L_{2+} = ty\phi$ follows from eq.~(5.9). Next, from eqs.~(5.9),(5.4) and (5.12), an explicit connection between current $r$ and $y$ is

$$
r = t^3(y' + y^2) + (a+1)t^2y.   \eqno(6.9)  
$$

\ni Recalling eq.~(6.5) and the fact that $L_{1a}^2 = L_{1+}L_{1-} = \phi L_{1-}$, we find the expression for $L_{1-}$:



$$
L_{1-} = \frac{h}{4t^2y\phi}, \ \ \ \ \ h \equiv \frac{r^2-1}{t^2y} - 2(a-1)(r+1).  \eqno(6.10)
$$

\ni From eq.~(5.10) we find 

$$
L_{2-} = -\frac{r^2-1}{4t^3y\phi}.   \eqno(6.11)
$$

\ni Finally the Lax matrices read:

$$
L_0 = \left(\begin{array}{cc}  0 & 0 \\ 0 & 1  \end{array}\right), \ \ \ \ \ L_1 = \left(\begin{array}{cc}  0 & \phi \\ \frac{h}{4t^2y\phi} & 1-a  \end{array}\right), \ \ \ \ \ L_2 = \left(\begin{array}{cc}  \frac{1+r}{2t} & ty\phi \\ -\frac{r^2 - 1}{4t^3y\phi} &  \frac{1 - r}{2t}  \end{array}\right),
$$

\ni and, using eqs.~(3.21), (3.22), $\hat B=0$ and (3.40), we get

$$
B_{-1} = \frac{1}{2t^2}\cdot I + \frac{1}{t}\L_2 = \left(\begin{array}{cc}  \frac{1+r}{2t^2} & y\phi \\ -\frac{r^2 - 1}{4t^4y\phi} & \frac{1-r}{2t^2}  \end{array}\right), \ \ \ \ \ B_1=0, \ \ \ \ \ B_0 = B_{0t}\cdot I = \left(\frac{1+r}{2t^2} - \frac{h}{4t^3y}\right)\cdot I.
$$


\ni Thus, the whole Lax pair reads:

$$
L =  \left(\begin{array}{cc} \frac{1+r}{2tx^2}  & \frac{\phi}{x} + \frac{ty\phi}{x^2} \\ \frac{h}{4t^2y\phi}\frac{1}{x} - \frac{r^2 - 1}{4t^3y\phi}\frac{1}{x^2} & \frac{1-r}{2tx^2} + \frac{1-a}{x} + 1  \end{array}\right),
$$

$$
B = \left(\frac{1+r}{2t^2} - \frac{h}{4t^3y}\right)\cdot I + \frac{1}{x}\left(\begin{array}{cc}  \frac{1+r}{2t^2} & y\phi \\ -\frac{r^2 - 1}{4t^4y\phi} & \frac{1-r}{2t^2}  \end{array}\right),    \eqno(6.12)
$$

\ni where $y=\phi'/\phi$ is a solution of Painlev\'e III eq.~(2.7a), $h$ is defined in eq.~(6.10) and $r$ is a solution of eq.~(6.6), which is related with $y$ by eq.~(6.9). Also, $y$ can be expressed in terms of $r$ from eq.~(5.17) (where $l_d=0$ now) as 

$$
y = \frac{r^2-1}{t^2(tr' + (a-1)(r+1))}.    \eqno(6.13)  
$$ 

\ni With $r$ replaced by $q$ according to eq.~(6.7), the above Lax pair will take exactly the form of the first Lax pair from Theorem 1.

\subsection{Case Ib} 

Also $\beta=2$ here. Since $L_{2+} = 0$, eq.~(4.2) gives $L_{2d}^2 = 1/(4t^2)$. To get a simpler solution, we choose $C_{2d}=-1$, i.e. $L_{2d} = -\frac{1}{2t}$. From eq.~(5.31) it follows then that  $L_{0+} = -t^2\phi' = -t^2y\phi$. The simplest solution is obtained if one takes $\hat B = \L_1$ in eqs.~(5.25)--(5.27). Then from eq.~(5.28) we find that $\phi'=y\phi$ and $A_- = A_1$. It is also the simplest to take $l_d \equiv -C_{2d}l_- = 0$, which solves eq.~(5.36) and means that $L_{2-} = 0$, by eqs.~(5.28) and (5.1). From eq.~(5.34) one finds then that $L_{1d} = \frac{a}{2}$. From eq.~(5.38) it follows now that $L_{0d}=r/2$, from eq.~(5.32) one finds



$$
L_{0-} = \frac{r^2 - 1}{4t^2y\phi},   \eqno(6.14)  
$$

\ni while eq.~(5.37), together with eqs.~(5.34) and (5.28), leads to 


$$
L_{1-} = \frac{A_1}{\phi} = \frac{h}{4t^2y\phi}, \ \ \ \ \ h \equiv \frac{r^2-1}{t^2y} + 2a(r+1).  \eqno(6.15)  
$$

\ni Now we recall that $B_1 = 0$ by eq.~(3.22), $B_{-1} = -\prt_tL_{2t}\cdot I + b_-(t)\L_2 = 1/(2t^2)\cdot I + \L_2/t$ by eq.~(3.21). Finally, from eq.~(3.40), recalling also that here $L_{1t} = C_{1t} = -a/2$, we find



$$
tB_{0t} = \frac{1+r}{2t} - \frac{h}{4t^2y} - \frac{a}{2}.    
$$

\ni Gathering all this information together, we can write down

$$
L_2 = \left(\begin{array}{cc}  0 & 0 \\ 0 & \frac{1}{t}  \end{array}\right), \ \ \ \ \ L_1 = \left(\begin{array}{cc}  0 & \phi \\ \frac{h}{4t^2y\phi} & -a  \end{array}\right), \ \ \ \ \ L_0 = \left(\begin{array}{cc}  \frac{1+r}{2} & -t^2y\phi \\ \frac{r^2 - 1}{4t^2y\phi} &  \frac{1 - r}{2}  \end{array}\right),
$$

$$
B_{-1} = \left(\begin{array}{cc}  0 & 0 \\ 0 & \frac{1}{t^2}  \end{array}\right), \ \ \ \ \ B_1=0,
$$

$$
B_0 = B_{0t}\cdot I + \frac{1}{t}\L_1 = \left(\frac{1+r}{2t^2} - \frac{h}{4t^3y} - \frac{a}{2t}\right)\cdot I  + \frac{1}{t}\left(\begin{array}{cc}  \frac{a}{2} & \phi \\ \frac{h}{4t^2y\phi} & -\frac{a}{2}  \end{array}\right),
$$

\ni and the whole Lax pair reads:

$$
L =  \left(\begin{array}{cc} \frac{1+r}{2}  & \frac{\phi}{x} - t^2y\phi \\ \frac{h}{4t^2y\phi}\frac{1}{x} + \frac{r^2 - 1}{4t^2y\phi} & \frac{1}{tx^2} - \frac{a}{x} + \frac{1 - r}{2}  \end{array}\right),
$$

$$
B = \left(\frac{1+r}{2t^2} - \frac{h}{4t^3y}\right)\cdot I + \left(\begin{array}{cc}  0 & \frac{\phi}{t} \\ \frac{h}{4t^3y\phi} & \frac{1}{t^2x} - \frac{a}{t}  \end{array}\right),   \eqno(6.16)
$$

\ni where $y=\phi'/\phi$ is a solution of Painlev\'e III eq.~(5.45) (with $C_{2d}=-1$), $h$ is given in eq.~(6.15), $y$ and $r$ are tied by the system, see eqs.~(5.39) and (5.40), where now $l_d=0$ and $A_-=A_1$ is given by eq.~(6.15),

$$
t(t^2y)' = r - t^3y^2 + at^2y,   \eqno(6.17)  
$$

$$
tr' = \frac{r^2-1}{t^2y} + a(r+1).   \eqno(6.18)  
$$

\ni It is easy to find from the above that $r$ itself satisfies the following second-order ODE:

$$
t(r^2-1)(tr')' = r(tr')^2 + \frac{1}{t}(r^2-1)^2 - a^2(r+1)^2.    \eqno(6.19)  
$$

\ni If we again introduce function $q$ as in eq.~(6.7), $r = -1 + 2q^2$, which is inspired by the hard-to-soft edge limit, see below, then the new $q$ will satisfy



$$
t(q^2-1)(tq')' = q(tq')^2 + \frac{q}{t}(q^2-1)^2 - \frac{a^2}{4}q,    
$$

\ni which is nothing but the Tracy-Widom equation (2.12) (or (5.67)). Then the above Lax pair will take exactly the form of the second Lax pair from Theorem 1.

\subsection{Case Ic} 

Here $\beta=4$, one has $L_{1d} = 0$, $S_d=tL_{2d}+L_{0d}=0$ and $A_a=tL_{2a}-L_{0a} =0$ by construction. Let $A_d = q$ -- the solution of hard edge Tracy-Widom eq.~(5.67), then immediately from eqs.~(5.50), (5.1) and the fact that $S_d=0$ we get

$$
L_{2d} = \frac{q}{2t}, \ \ \ \ \ L_{0d} = -\frac{q}{2}.   \eqno(6.20)  
$$

\ni One has also $S_a=tL_{2a}+L_{0a} = 2tL_{2a}=2L_{0a}$. Eq.~(5.56) yields $S_a^2 = S_+S_- = 1-q^2$, while eq.~(5.65) gives $L_{1a}^2 = L_{1+}L_{1-} = \frac{4(tq')^2-a^2}{4(q^2-1)}$. From eqs.~(5.57) and (5.60) one finds







$$
2S_+L_{1-} = 2tq'-a, \ \ \ \ \ 2S_-L_{1+} = -2tq'-a,   \eqno(6.21)  
$$

\ni which, together with eqs.~(5.56) and (5.65) gives also the ratios

$$
\frac{L_{1+}}{S_+} = \frac{2tq'+a}{2(q^2-1)}, \ \ \ \ \ \frac{L_{1-}}{S_-} = -\frac{2tq'-a}{2(q^2-1)}.   \eqno(6.22)  
$$

\ni We recall also the eqs.~(5.58) and (5.59) which mean, respectively, that

$$
tS_+' = qL_{1+}, \ \ \ \ \ tS_-' = -qL_{1-},   \eqno(6.23)  
$$

$$
t^2L_{1+}' = qS_+, \ \ \ \ \ t^2L_{1-}' = -qS_-.  \eqno(6.24)  
$$

\ni From eqs.~(6.22) and (6.23) we express the logarithmic derivatives of $S_+$ and $S_-$,

$$
\frac{S_+'}{S_+} = \frac{qq'}{q^2-1} + \frac{aq}{2t(q^2-1)}, \ \ \ \ \ \frac{S_-'}{S_-} = \frac{qq'}{q^2-1} - \frac{aq}{2t(q^2-1)},    \eqno(6.25)  
$$

\ni and the components $L_{1+}$ and $L_{1-}$ are then expressed from eq.~(6.22), which completes finding of the components of matrix $L$ in terms of the solution $q$ of eq.~(5.67). Recalling also that $L_{2t} = 1/(2t)$, $L_{1t} = -a/2$ and $L_{0t}=1/2$, we can write out

$$
L_2 = \left(\begin{array}{cc}  \frac{1+q}{2t} & \frac{S_+}{2t} \\ \frac{S_-}{2t} & \frac{1-q}{2t}  \end{array}\right), \ \ \ \ \ L_1 = \left(\begin{array}{cc} -\frac{a}{2} & \frac{2tq'+a}{2(q^2-1)}S_+ \\ -\frac{2tq'-a}{2(q^2-1)}S_- & -\frac{a}{2}  \end{array}\right), \ \ \ \ \ L_0 = \left(\begin{array}{cc}  \frac{1-q}{2} & \frac{S_+}{2} \\ \frac{S_-}{2} & \frac{1+q}{2}  \end{array}\right),
$$



\ni where the functions $S_+$ and $S_-$ are completely determined by eqs.~(2.13). Now we recall that $B_1=0$ by eq.~(3.22), $B_{-1} = 1/(2t^2)\cdot I + \L_2/t$ by eq.~(3.21), $\B_0 = \L_1/(2t)$ by eqs.~(3.41) and $\hat B = \L_1$. Finally, by eq.~(3.40),




$$
2tB_{0t} = -\frac{(2tq')^2-a^2}{4(q^2-1)} + \frac{q^2}{t} + \frac{a(a-2)}{4}.
$$

\ni Thus, after gathering all this together, the Lax pair for $\beta=4$ is the one of Theorem 2.






\subsection{Case II}

This is a case where $\kappa = 2-a$, induced by requiring that $L_{2t} = 1/(2t)$ and $L_{0t} \neq 1/2$. The final result depends on two arbitrary functions, which we chose in such a way that the Lax pair bears some resemblance with the previous Lax pairs where all the functions are determined, unlike here.
\par Again let $X_2=t\L_2$ as in eq.~(5.1). One of the components of eq.~(3.42) is redundant due to eq.~(4.2) being its first integral. The other two components depend on two arbitrary functions as e.g.

$$
\kappa tX_{2d}' = -\hat b_-X_{2+}, \ \ \ \ \ \kappa tX_{2+}' = 2\hat b_dX_{2+},
$$

\ni where, using eqs.~(4.17) and (4.23) we represented the matrix $\hat B$ from eq.~(3.41) as 

$$
\hat B = \frac{1}{t}X_2 + \hat b_d\sigma_3 + \hat b_-\sigma_-.
$$

\ni We denote the two arbitray components as $X_{2d}=r/2$ and $X_{2+}=\phi$. Then eq.~(4.2) gives $X_{2-} = -\frac{r^2-1}{4\phi}$, and the other components of matrix $L$ are already known from section 4. Using eqs.~(3.21), (3.22) and (3.40) we find the non-arbitrary elements of matrix $B$. The final result is the Lax pair of Theorem 3.



\par One can easily verify that, for {\it any} two functions $r(t)$ and $\phi(t)$, the zero curvature equations are satisfied {\it identically} with this pair. The Fokker-Planck equation is satisfied by the first component $\F^H$ of the eigenvector of the pair provided that $\kappa = 2-a$ only, but with no other restrictions. However, the above Lax pair means that 



$$
\prt_x\F^H = \frac{1}{2tx^2}\left((1+r)\F^H + 2\phi G\right)
$$

\ni and

$$
\prt_t\F^H = \frac{1}{2t^2}\left(\frac{1}{x}+\frac{1}{\kappa}\right)\left((1+r)\F^H + 2\phi G\right),
$$

\ni which are multiples of each other, and so this $\F^H$ in fact also satisfies a much simpler equation,

$$
\kappa t\prt_t\F^H = x(x+\kappa)\prt_x\F^H,
$$

\ni and, as a consequence of this and the FP eq.~(2.2), also satisfies

$$
\left(\prt_x + \frac{2}{x} - \frac{1}{tx^2}\right)\prt_x\F^H = 0,
$$

\ni both of which can be solved and with proper initial conditions for $\F^H$ result in eqs.~(2.16) and (2.17) of Theorem 3 for $\F^H$ and $G$.




\section{Hard edge to soft edge limit transition}

This transition was considered before in~\cite{BorFor} in terms of Fredholm determinants and its kernels and in terms of stochastic 1st order ODEs in~\cite{RR08}. Here we track how it goes in terms of the derived above Painlev\'e equations, which partly complements considerations of~\cite{ForPainl00}, and related Lax pairs, for the cases $\beta=2$ and $4$. The transition occurs in the limit as $a\to\infty$, then, if we denote $\alpha = \frac{a}{2}$, the independent variables $x$ and $t$ in the FP equation (2.2) scale as in eq.~(2.18), i.e.~$x \to \alpha(1+\alpha^{-1/3}x)$, $t \to \alpha^{-2}(1+\alpha^{-2/3}t)$. Then, since the partial derivatives scale as 



$$
\prt_x \to \alpha^{-2/3}\prt_x,  \ \ \ \ \ \prt_t \to \alpha^{8/3}\prt_t,   \eqno(7.1)  
$$

\ni the hard edge FP operator $\D_H$ scales as 


$$
\D_H = (\kappa t\prt_t + x^2\prt_{xx} + (ax - x^2 - 1/t)\prt_x) \to \alpha^{2/3}\D_S = \alpha^{2/3}(\kappa\prt_t + \prt_{xx} + (t-x^2)\prt_x),
$$



\ni where we have the soft edge FP operator written out by~\cite{BV} on the right-hand side. This simply and clearly justifies the name of the transition, and shows also that the solution of eq.~(2.2), $\F^H$, scales as $\F^H \to \alpha^{-2/3}\F^S$, with $\F^S$ being the solution of soft edge FP equation~\cite{BV}.



\subsection{The transition for $\beta=2$}
 
We only write down the derivation for the Case Ia of Lax pair (2.6a) since the transition for Case IIb of pair (2.6b) gives the same final result (even though the scaling of functios involved is different which can be most easily figured out from the algebraic identities presented in the remark after Theorem 1, e.g.~the scaling of $y$ changes to $y\to \alpha^3(1+\alpha^{-1/3}y)$ instead of that in eq.~(7.15) below and $G$ corresponding to eq.~(7.19) below changes to $G \to e^{\alpha^{-1/3}t}G^S$, however, the scaling of $q$ does not change). 
Recall the defining system of first-order ODE eqs.~(6.9) and (6.13) obtained in section 6.1 for functions denoted $y$ and $r$ ($y$ being the solution of Painlev\'e III eq.~(2.7a)):




$$
t^3y' = r - t^3y^2 - (a+1)t^2y,   \eqno(7.2)  
$$

$$
t^3yr' = (r+1)\left(r-1 - (a-1)t^2y\right).   \eqno(7.3)  
$$

\ni We consider the scaling limit of these equations, using eqs.~(2.18) and (7.1) and substituting

$$
y \to y_0(\alpha) + \alpha^{\Delta_y}y_1, \ \ \ \ \  r \to r_0(\alpha) + \alpha^{\Delta_r}r_1.   \eqno(7.4)  
$$

\ni Then the highest order in $\alpha$ gives, repectively,

$$
 r_0 - 2\alpha^{-3}y_0 - \alpha^{-6}y_0^2 = 0,    \ \ \ \ \ (r_0+1)(r_0-1 - 2\alpha^{-3}y_0) = 0,
 $$
 
 \ni with possible solutions $r_0=-1, \ y_0 = -\alpha^3$ and $r_0=3, \ y_0=\alpha^3$. Taking the next terms in $\alpha$ shows that only the first solution is consistent. Then the next order terms in (7.2) give 
 
 
 
 $$
  \alpha^{\Delta_y-10/3}y_1' = \alpha^{\Delta_r}r_1 + \alpha^{-2/3}t - \alpha^{2(\Delta_y-3)}y_1^2.  \eqno(7.5)  
 $$
 
 \ni Similarly, the next order terms in eq.~(7.3) yield 
 
 
 
 $$
 \alpha^{-1/3}r_1' = r_1(2\alpha^{\Delta_y-3}y_1 - \alpha^{\Delta_r}r_1).   \eqno(7.6)  
 $$
 
 \ni Equations (7.5) and (7.6) consistently imply that the scaling dimensions are
 
 $$
\Delta_y = \frac{8}{3}, \ \ \ \ \ \Delta_r = -\frac{2}{3},   \eqno(7.7)  
$$

\ni and thus, eqs.~(7.5) and (7.6) finally become the corresponding system for the soft edge limit:

$$
y_1' = r_1 - y_1^2 +t,   \eqno(7.8)  
$$

$$
r_1' = 2y_1r_1.   \eqno(7.9)  
$$

\ni Solving it for $y_1$ and $r_1$, respectively, one gets

$$
y_1'' = 2y_1^3 - 2ty_1 -1   \eqno(7.10)  
$$

\ni and

$$
2r_1r_1'' = (r_1')^2 + 4tr_1^2 + 4r_1^3.   \eqno(7.11)  
$$

\ni Equations (7.10) and (7.11) can, of course, also be obtained by taking the scaling limit of the corresponding second-order ODEs eq.~(2.7a) and eq.~(6.6). The transformation

$$
y_1 = 2^{1/3}y, \ \ \ \ \ t = -s/2^{1/3}
$$

\ni brings eq.~(7.12) to the standard Painlev\'e II form:

$$
y''(s) = sy + 2y^3 - \frac{1}{2},   \eqno(7.12)  
$$

\ni while (7.11) is a form of Painlev\'e XXXIV equation, see e.g.~\cite{Clarkson2011}. If one substitutes

$$
r_1 = 2q^2,   \eqno(7.13)  
$$

\ni then one obtains the Painlev\'e II equation with free parameter zero:

$$
q''(t) = tq + 2q^3,   \eqno(7.14) 
$$

\ni which has the Hastings-McLeod solution relevant to the soft edge limit~\cite{TW-Airy}. It may seem surprising that another Painlev\'e II, with free parameter $-1/2$ also arises here. This, however, has a natural explanation in the fact that the solutions of the two Painlev\'e II equations are tied by the Gambier relation, see e.g.~recent nice Painlev\'e II review~\cite{ForWit2012}:

$$
2^{1/3}\epsilon q_0^2(-2^{-1/3}t) = \frac{d}{dt}q_{\epsilon/2}(t) + \epsilon q_{\epsilon/2}^2(t) + \frac{\epsilon}{2}t,   
$$

\ni where $q_{\nu}$ means a solution of Painlev\'e II in the standard form with free parameter $\nu$, and $\epsilon=\pm 1$. This relation also naturally explains the discrepancy of independent variables in eqs.~(7.12) and (7.14).
\par Now let us consider the soft edge limit for the hard edge Lax pair (6.12) of subsection 6.1 (this is the Lax pair (2.6a) in terms of $r$ instead of $q$ from eq.~(2.12)).





\par One should have $L \to \alpha^{-2/3}L_S$ and $B \to \alpha^{8/3}B_S$, according to eq.~(7.2), where $L_S$ and $B_S$ are the limiting soft edge matrices to be found. One finds, however, with the help of eqs.~(2.18), (7.1) and

$$
y \to -\alpha^3(1-\alpha^{-1/3}y),   \ \ \ \ \ r \to -1 + \alpha^{-2/3}r,   \eqno(7.15)  
$$

\ni that

$$
L \to \alpha^{-2/3} \left(\begin{array}{cc} \frac{r}{2}  & \alpha^{-2/3}e^{-\alpha^{1/3}t}(x + y)\phi \\ \alpha^{2/3}e^{\alpha^{1/3}t}\cdot\frac{r(x - y)}{2\phi} & x^2-t-\frac{r}{2}  \end{array}\right),   \eqno(7.16)  
$$

$$
B \to \alpha^{8/3}\frac{r}{2}\left(y^2-t-\frac{r}{2}\right)\cdot I + \alpha^{8/3} \left(\begin{array}{cc} \alpha^{-1/3}\frac{r}{2}  & -e^{-\alpha^{1/3}t}\phi \\ -e^{\alpha^{1/3}t}\cdot\frac{r}{2\phi} & \alpha^{1/3} - x  \end{array}\right),   \eqno(7.17) 
$$

\ni the exponential factors appear because $\phi$ scales as 

$$
\phi \to e^{-\alpha^{1/3}t}\phi,     \eqno(7.18) 
$$

\ni and new $\phi$ again satisfies the relation $\phi' = y\phi$ with the new $y$ from (7.15). Rescaling $\F^H$ according to (2.22) removes the additional $\alpha^{\pm2/3}$ factors in the non-diagonal entries of $L$. The exponential factors $e^{\pm\alpha^{1/3}t}$ as well as the term $\alpha^{1/3}$ in the last entry of $B$ in (7.17) cancel upon rescaling of the second component of the eigenvector $G$ as

$$
G \to e^{\alpha^{1/3}t}G^S,   \eqno(7.19)  
$$

\ni which determines the correct scaling of $G$ in the transition to the soft edge. Thus, finally we obtain the corresponding soft edge limit Lax pair:

$$
\prt_x\left(\begin{array}{c}\F^S \\ G^S \end{array}\right) = L_S\left(\begin{array}{c}\F^S \\ G^S \end{array}\right),  \ \ \ \ \ \prt_t\left(\begin{array}{c}\F^S \\ G^S \end{array}\right) = B_S\left(\begin{array}{c}\F^S \\ G^S \end{array}\right),  \eqno(7.20) 
$$

\ni where

$$
L_S = \left(\begin{array}{cc} q^2  & (x + y)\phi \\ \frac{q^2(x - y)}{\phi} & x^2-t-q^2  \end{array}\right),   \eqno(7.21)  
$$

$$
B_S = \left(\begin{array}{cc} q^2(y^2-t-q^2)  & -\phi \\ -\frac{q^2}{\phi} & q^2(y^2-t-q^2) - x  \end{array}\right),   \eqno(7.22)  
$$
 
\ni and $q$ is the Hastings-McLeod solution of the Painlev\'e II equation (7.14) and $r = 2q^2$ for the rescaled $r$ on the right-hand side of the second relation in (7.15), see eq.~(7.13).  Now we infer from eqs.~(7.9) and (7.13) that our $\phi$ turns into $q$ and we obtain the Lax pair of Theorem 4.
 
\subsection{The transition for $\beta=4$}
 
Here it is convenient to consider the scaling limit of the hard edge Tracy-Widom equation (2.12). It turns out that under the scaling



$$
q \to \alpha^{-1/3}q   \eqno(7.23)  
$$

\ni and the scalings from (2.18) and (7.1) for $t$ and $\prt_t$ this equation turns exactly into the Painlev\'e II eq.~(7.14) with free parameter zero, $q'' = tq+2q^3$. Thus, eq.~(7.23) gives the right scaling for $q$, and then equations (2.13) determine that functions $S_+$ and $S_-$ scale as $\sim 1$ in $\alpha$. Thus, in the limit we indeed get eq.~(2.24). 




\par Consider now the limit for the Lax pair involving the above functions. This time $L$ rescales correctly from the beginning:

$$
L \to \alpha^{-2/3} \left(\begin{array}{cc} \frac{x^2-t}{2}-qx  & \left(\frac{x^2-t}{2} - q' - q^2\right)S_+ \\ \left(\frac{x^2-t}{2} + q' - q^2\right)\frac{1}{S_+} & \frac{x^2-t}{2}+qx   \end{array}\right),   \eqno(7.24)  
$$

\ni and so does $B$:

$$
B \to -\alpha^{8/3}\frac{1}{2}\left(\alpha^{1/3} - u \right)\cdot I + \alpha^{8/3} \left(\begin{array}{cc} \frac{\alpha^{1/3}}{2}-\frac{x-q}{2}  & -\frac{S_+x}{2} \\ -\frac{x}{2S_+} & \frac{\alpha^{1/3}}{2}-\frac{x+q}{2}  \end{array}\right) =
$$

$$
= \alpha^{8/3} \left(\begin{array}{cc} -\frac{x-u-q}{2}  & -\frac{S_+x}{2} \\ -\frac{x}{2S_+} & -\frac{x-u+q}{2}  \end{array}\right),   \eqno(7.25)  
$$

\ni where $u$ is the function satisfying eqs.~(2.21) in terms of the new $q$ satisfying eq.~(7.14). Therefore we infer also that here both of the eigenvector components rescale as in eq.~(2.25). 




\section{Conclusions}

We established the correspondence of solutions to FP equation for hard-edge probabilities of spiked large $\beta$-Wishart ($\beta$-Laguerre) matrices for special values of $\beta = 2,4$ with eigenvectors of certain Lax pairs for Painlev\'e III. Also it is shown that there is no such simple correspondence for other values of $\beta$, as is the case with the FP equation for the corresponding soft-edge probabilities~\cite{BV}. The only difference is the case when a special relation eq.~(2.1) holds between the additional parameter $a$ here and $\beta$. Then the bounded solution of the hard-edge FP equation is explicit elementary function, eq.~(2.16), which can be seen as a one-parameter generalization of the Gumbel distribution for extreme values.
\par Thus, the relation of the FP solutions, which represent non-stationary euclidean time eigenfunctions for quantum Painlev\'e Hamiltonians, with $\beta$ playing the role of the Planck constant (or its inverse), see~\cite{Nag11}, with the classical Painlev\'e equations and classical integrability in general remains an interesting open problem, further considered in~\cite{BetaPhil}.









\end{document}